\documentclass[aps,prñ,10pt,twocolumn,superscriptaddress,showpacs,floatfix]{revtex4-1}
\usepackage{graphicx}
\usepackage{dcolumn}
\usepackage{bm}
\usepackage{longtable}

\begin{document}

\title{Curvature correction term as a constraint for the Skyrme interaction}
\author{K. V. Cherevko}
\email{k.cherevko@univ.kiev.ua}
  \affiliation{College of Nuclear Science and Technology, Beijing Normal University, Beijing 100875, China}
  \affiliation{Beijing Radiation Center, Beijing 100875, China}
  \affiliation{Physics Faculty, Taras Shevchenko National University of Kyiv, 4 Glushkova av., Kyiv, 03022, Ukraine}
\author{L. A. Bulavin}
  \affiliation{Physics Faculty, Taras Shevchenko National University of Kyiv, 4 Glushkova av., Kyiv, 03022, Ukraine}
\author{L. L. Jenkovszky}
\email{jenk@bitp.kiev.ua}
  \affiliation{Bogolyubov Institute for Theoretical Physics (BITP), Ukrainian National Academy of Sciences, Metrolohichna str. 14-b, Kyiv, 03680, Ukraine}
\author{V. M. Sysoev}
  \affiliation{Physics Faculty, Taras Shevchenko National University of Kyiv, 4 Glushkova av., Kyiv, 03022, Ukraine}
\author{Feng-Shou Zhang}
 \email{fszhang@bnu.edu.cn}
  \affiliation{College of Nuclear Science and Technology, Beijing Normal University, Beijing 100875, China}
  \affiliation{Beijing Radiation Center, Beijing 100875, China}
  \affiliation{Center of Theoretical Nuclear Physics, National Laboratory of Heavy Ion Accelerator of Lanzhou, Lanzhou 730000, China}

\date{\today}

\begin{abstract}
The curvature correction term to the surface tension is used as a criterion for the efficiency of the Skyrme interaction in describing surface properties. Based on the nuclear equation of state, the curvature correction term to the surface tension coefficient is calculated for 97 standard Skyrme interaction parameter sets in the vicinity of nuclear saturation density at zero temperature. The main idea is to find those parametrizations that give the Tolman's $\delta$ correction close to the available theoretical predictions from the statistical theory. Only 59 out of 97 models give satisfactory results. Comparison of the obtained results with the results of the implementation of different macroscopic and microscopic constraints to Skyrme parametrizations available in the literature allows us to select 4 models that satisfy all the constraints.
\end{abstract}\texttt{}


\maketitle

\section{Introduction.}
The derivation of the equation of state (EOS) of nuclear matter is among the most important goals and long-standing unsolved problems in nuclear physics and astrophysics \cite{Baldo2007}. Various  approaches to the description of infinite nuclear matter exist. Among them are purely microscopic ones based on the realistic description of the nucleon-nucleon (NN) interaction \cite{Machleidt2001}, in which case the result depends not only on the chosen interaction but also on the way many-body effects are treated. These may be handled either by a direct description of the tree-particle interactions or by approaches like that of the Bruckner-Hartree-Fock method, the Dirac-Bruckner-Hartree-Fock formalism \cite{Haar1987, Bombaci1991} or self-consistent Green's functions \cite{Dewulf2003}, etc.

At the same time, in describing the experimental data and in computer simulation, the most widely used models are those based on effective density-dependent NN and NNN interactions rather than on realistic ones (e.g., the models introduced by Skyrme \cite{Skyrme1956} and Gogny \cite{Decharge1980}). The main problem of such approaches is in the infinite number of possible sets of model parameters providing satisfactory description of the ground-state properties of stable nuclei.

A large number of various Skyrme-force parametrizations and theoretical models attempting to describe nuclear matter and finite nuclei in a wide range of external parameters exists. They all were constructed under specific assumptions that reduce their predictive power \cite{Stevenson2012}. Combined with the indirect model-dependent experimental methods used to evaluate nuclear matter properties, it makes the selection of the realistic sets of parameters
quite a difficult task.

Presently over 200 sets of Skyrme parameters are known from the literature.
They result from the analysis of various observables, leading to different predictions concerning the behavior of nuclear matter away from equilibrium. Recently, a number of interesting and important papers systematically checking the sets of parameters for nuclear matter constraints appeared \cite{Stevenson2012, Dutra2012}. Important work on the construction of new parametrizations with systematic variation of the parameters to improve the precision of the results for some crucial nuclear matter properties has been done in Refs. \cite{Friedrich1986, Gomez1992, Klupfel2009}. Such investigations may result in the improvement of the equation of state of nuclear matter, applicable in a wide range of parameters. Thus, the search for model-independent constraints connected with the specific properties of the nuclear matter is timely and important.

Such constraints may result from the properties of the interfaces. To start with, let us mention that the Droplet model of nuclei \cite{Myers1969} plays a special role among  macroscopic models. It makes possible the description of average properties of a saturated system, such as a nucleus, consisting of two components (neutrons and protons), with account for the boundary effects and the presence of a diffuse layer. The surface energy and the properties of the surfaces in nuclear matter have been studied in a number of papers \cite{Ravenhall1983, Boyko1990, Jenkovszky1994}. Although the dependence of the surface tension (and surface energy) on the surface curvature as well as its impact on different physical properties were also studied by several groups of authors \cite{Moretto2012, Pomorski2003}. Still, for decades it remains one of the most controversial issues in mesoscopic thermodynamics \cite{Anisimov2007, Blokhius2006, Kolomietz2012}.

In studies of surface properties of nuclei with mass number $A$, the account for the curvature effects is important. Within the Droplet model this requires the inclusion of additional terms proportional to $A^{\frac{1}{3}}$ in any expansion concerning the nuclear properties in terms of the fundamental dimensionless ratio, given by $\frac{r_0}{R}=A^{-\frac{1}{3}}$, which is the ratio of the interparticle spacing $r_0$ to the nuclear radius $R$ \cite{Myers1969}. In these studies, effects connected with the surface curvature are specified by the curvature correction coefficient $a_3$ accompanying terms of the order $A^{\frac{1}{3}}$. In statistical mechanics and in Gibbs-Tolman's (G-B) thermodynamics,
of interfaces it corresponds to to the Tolman length, called also $\delta-$ correction \cite{Ono1960}. The basic parameter $\delta$ was first introduced by Tolman  \cite{Tolman1949}. It is equal to the distance between the equimolar surface $R_{em}$ and the surface of tension $R$ at the interphase boundary
\begin{equation}
\label{delt1}
   \delta=R_{em}-R.
 \end{equation}
According to the G-T theory, the surface tension $\sigma$ of the curved interface, in the leading order approximation, can be defined as
\begin{equation}
\label{delt2}
   \sigma(R)=\sigma_\infty\left(1-\frac{2\delta}{R}+\cdot\cdot\cdot\right),
 \end{equation}
where $R$ is the droplet radius (equal to the radius of the surface of tension \cite{Rowlinson1982, Rowlinson1994}) and $\sigma_\infty$ is the surface tension of the planar interface.
Originally introduced for ordinary liquids, it can be defined for any system with curved interface of a non-negligible boundary layer \cite{Anisimov2007}, such as nuclei and nuclear systems with a finite diffuse layer \cite{Brack1985}.

First theoretical estimates of the correction term were done by Tolman \cite{Tolman1949a}. It appeared to be close to the average interparticle distance $r_0$. Namely, $\delta\sim0.3-0.6r_0$ , that, for a nuclear systems, is $r_0\sim0.7$ fm at normal density $\rho\sim0.17$ fm$^{-3}$.
Present calculations from statistical mechanics yield Tolman length of the order of the interparticle distance $\delta\sim{r_0}=1.14$fm \cite{Jenkovszky1994}. Thus, mathematically the term $\frac{2\delta}{R}$ in Eq. (\ref{delt2}) becomes important for the systems with $R<14$ fm ($\frac{2\delta}{R}>0.1$ in (\ref{delt2}) and even more so for nuclear systems with $R\approx0.7A^{\frac{1}{3}}=0.7(277)^{\frac{1}{3}}=<4.6$ fm for heavy nuclei.

In view of the importance of the curvature correction for nuclear systems, checking different sets of standard Skyrme parameters as for their ability to reproduce the theoretically predicted values for $\delta$-correction becomes an important task. Therefore, following the idea of \cite{Dutra2012} in this work we present an attempt to use Tolman correction as a constraint for different sets of Skyrme parameters.

\section{Theoretical Model.}
In the present paper the Tolman $\delta$-corrections are calculated for 97 different sets of Skyrme parameters known from the literature (see Appendix \ref{appA}). In our analysis we included the most popular parametrizations. From large families (e.g. BSk,SkSC), we selected several representative members. The Tolman length was calculated for the whole families in which $\delta$ coincides with or is close to that calculated statistically. The results are compared with the theoretical predictions.
This analysis is interesting as a test of various models regarding their ability to describe interphase interfaces.

Various approaches for the evaluation of the curvature correction exist. A method to calculate the Tolman $\delta$-correction from the EOS of nuclear mater was introduced earlier in \cite{Cherevko2014a}.

In that model, one gets for $\delta$ from the EOS of symmetric nuclear matter with isospin-independent effective mass (see Appendix \ref{appB}) in the case of $T=0,$ at normal density $\rho_0$:
\begin{equation}
\label{cor1}
 \begin{array}{l}
   \delta=\frac{2}{3}\frac{1}{{\rho_0}^2}\\
   \times\frac{-33t_0-160W{\rho_0}^{-1/3}+t_3(1+\alpha){\rho_0}^\alpha\frac{1}{12}\left(7(3\alpha+6)-3(3\alpha+6)^2\right)}{\left(15t_0+\frac{1}{12}t_3(1+\alpha)\left((3\alpha+6)-(3\alpha+6)^2\right)\right)^2}\sigma_\infty,
 \end{array}
\end{equation}
where
\begin{equation}
\label{cor2}
   W=\frac{h^2}{10m}\left(\frac{3}{8{\pi}g}\right)^\frac{2}{3}\left(\frac{5-3\frac{m*}{m}}{\frac{m*}{m}}\right).
   \end{equation}

\begin{figure}[htp]
\begin{minipage}{\linewidth}
\includegraphics[scale=1.0]{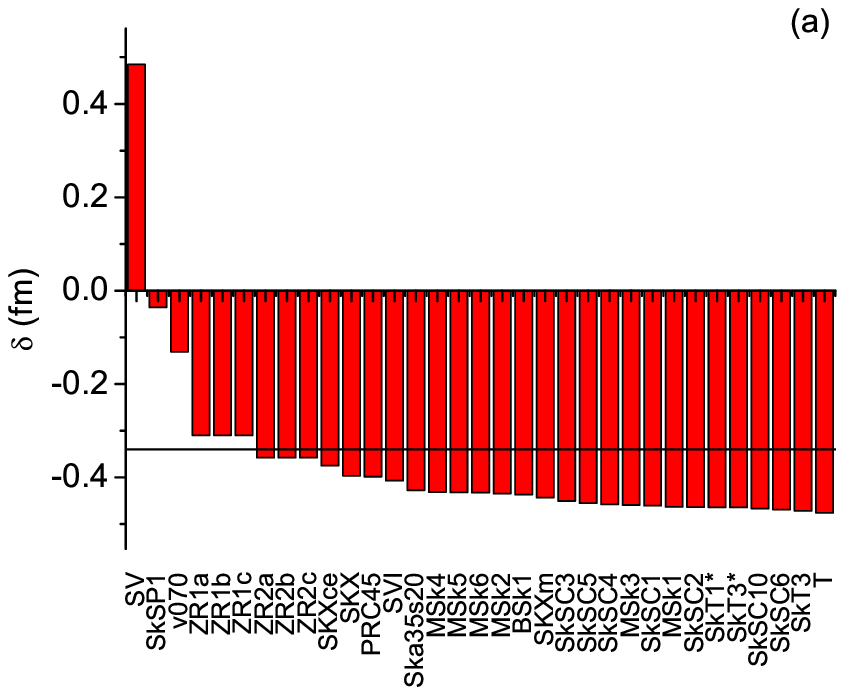}
\end{minipage}
\vfill
\begin{minipage}{\linewidth}
\includegraphics[scale=1.0]{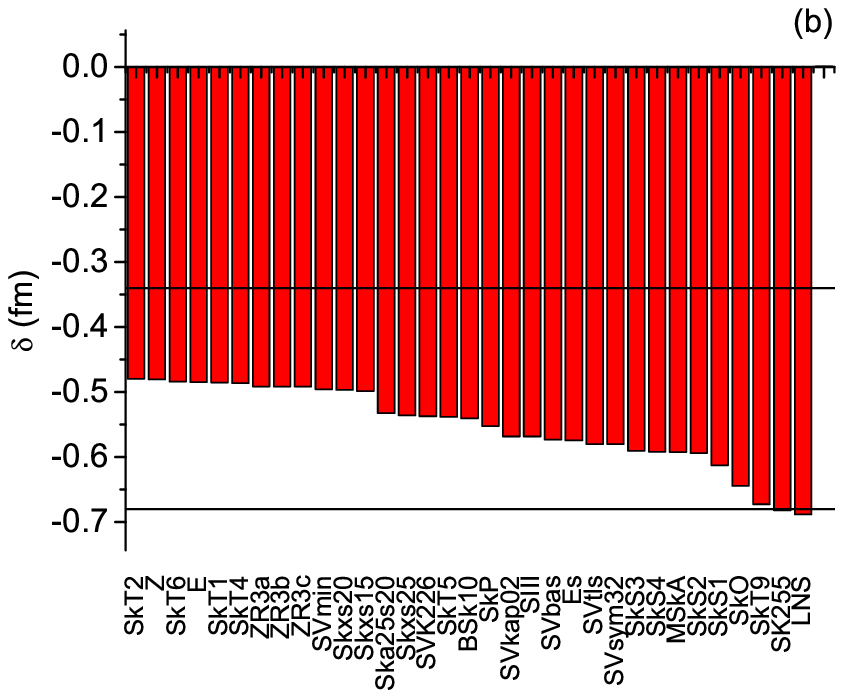}
\end{minipage}
\vfill
\begin{minipage}{\linewidth}
\includegraphics[scale=1.0]{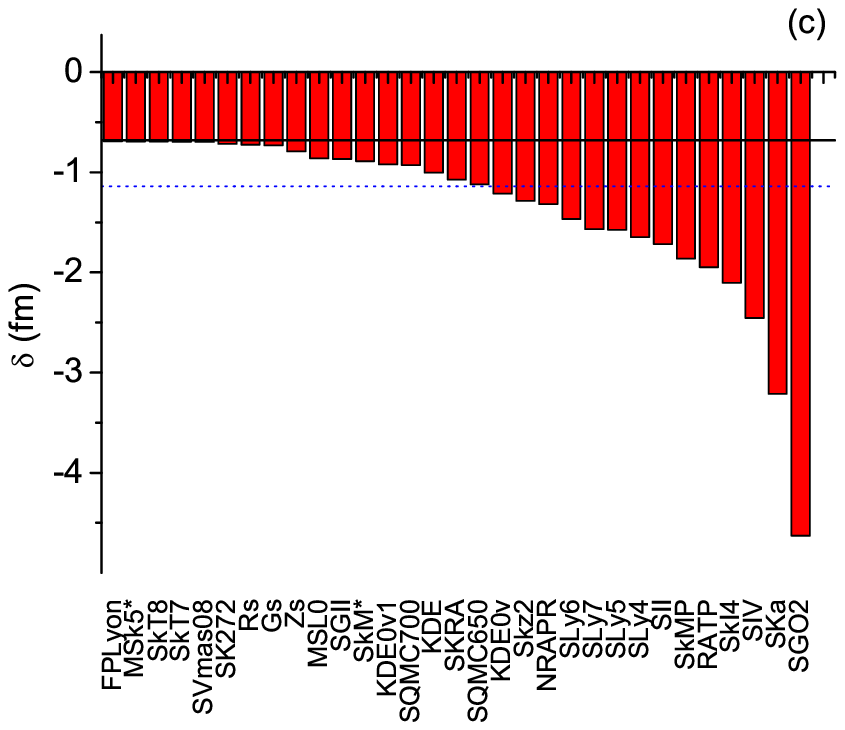}
\end{minipage}
\caption{Color online. Tolman's $\delta$-length. Solid lines show the boundaries defined by (\ref{res}); dashed line correspond to the extended range up to interparticle distance}
\label{fdelt}
\end{figure}

We use the above results to calculate corrections to 97 sets of Skyrme parameters, using them in testing various parametrizations describing surface effects properly. Let us recall that, following Tolman's first estimates
\cite{Tolman1949a}, subsequent calculations based on statistical mechanicsthat use minimum information regarding the details of the interaction about the studied liquids but operating with dimensionless parameters such as the size of the diffuse layer divided by the interparticle distance (e.g. penetrable sphere model in the mean field approximation)yield $\mid\delta\mid$ in a range from $\frac{1}{3}r_0$ to $\frac{5}{9}r_0$ (\cite{Harasima1958, Rowlinson1982, Gorski1989, Irving1950} and references therein). Even though the sign of the curvature corrections reported in the literature varies in different calculations, the absolute value is approximately the same, varying within the distance where the density profile faces rapid changes. The density profiles in the diffuse layer in the dimensionless coordinates are very similar for ordinary liquids and nuclear matter. This allows us to extend the results of the ordinary liquids physics to nuclear matter. Thus, with account for the uncertainty of the theoretical values of $\delta$, the admissible range for the curvature corrections adopted in the work is
\begin{equation}
\label{res}
   \mid\delta\mid=(0.3\sim0.6)r_0=(0.34\sim0.68) fm.
\end{equation}
In all calculations the surface tension of the semi-infinite matter at $T=0$ is calculated for the symmetric case with $\rho_n=\rho_p=\frac{1}{2}\rho$ (without Coulomb interaction) within the restricted extended Thomas-Fermi (ETF) approach \cite{Brack1985}. Terms up to fourth order are considered. All the calculations include effective mass and spin-orbit contribution. The function $\rho(z)$ used to minimize the surface energy is the one parameter Fermi function:
\begin{equation}
\label{st}
   \rho(z)=\frac{\rho_\infty}{1+exp({\alpha}z)}
\end{equation}
Our calculations yield values for $\sigma_\infty$ in the range $0.93\div1.21 MeV{\cdot}fm^{-2}$.
The results of our calculation are presented in Fig. \ref{fdelt} .

In Table \ref{result} the parametrization yielding values of $\delta$ within the admissible range (\ref{res}) and, at the same time, satisfy the constraints of Ref. \cite{Dutra2012} are shown, appended by the macroscopic properties of the corresponding parametrizations. The label "part" indicates that the parametrization passed the test of Ref. \cite{Dutra2012} for all but one of the applied constraints, and the failure for that one constraint is less than 5 per cent.
\begin{table}[htp]
\caption{Tolman $\delta$-correction and macroscopic properties for various Skyrme parametrizations. $\delta$ is in fm; $\rho_0$ is fm$^{-3}$; $K$ is in MeV; ${C_0}^\rho$ is in MeV fm$^3$. Forces below "Extended" pass the $\delta$ - correction test with the extended range (\ref{res}) up to interparticle distance}
 \label{result}
 \begin{ruledtabular}
 \begin{tabular}{lcccccc}
 \textrm{Skyrme force} & \textrm{$\delta$}& \textrm{$\rho_0$} & \textrm{$K$}& \textrm{$E_0$} & \textrm{${C_0}^\rho$}&\textrm{\cite{Dutra2012} results}\\*[3pt]
 Ska25s20 & -0.74 & 0.1746 & 210.78 & -15.32 & -250.23 & +\\
 SV-min & -0.74 & 0.1746 & 210.78 & -15.32 & -250.23 & +\\
 Ska35s20 & -0.78 & 0.16 & 230 & -16 & -272.70 & part\\
 SkT1 & -0.58 & 0.162 & 201.95 & -15.81 & -240.38 & part\\
 SkT2 & -0.58 & 0.162 & 201.95 & -15.81 & -240.38 & part\\
 SkT3 & -0.58 & 0.162 & 201.95 & -15.81 & -240.38 & part\\
 Skxs20 & -0.58 & 0.1595 & 234 & -15.94 & -253.67 & part\\
 \textbf{Extended} & $\;$ & $\;$ & $\;$ & $\;$ & $\;$ & $\;$\\
 LNS & -0.74 & 0.1746 & 210.78 & -15.32 & -250.23 & +\\
 SQMC700 & -0.74 & 0.1746 & 210.78 & -15.32 & -250.23 & +\\
 MSL0 & -0.58 & 0.1595 & 234 & -15.94 & -253.67 & part\\
 SKRA & -0.58 & 0.1595 & 234 & -15.94 & -253.67 & part\\
 KDE0v1 & -0.58 & 0.1595 & 234 & -15.94 & -253.67 & part\\
 \end{tabular}
\end{ruledtabular}
\end{table}

\section{Results and Discussion.}
Only 59 parametrization of those analyzed satisfy the constraint on the $\delta-$corrections imposed in Eq. (\ref{res}). Among them \textbf{Ska25s20} and \textbf{SV-min0}satisfy also all the criteria of Ref. \cite{Dutra2012}, while \textbf{Ska35s20,SkT1, SkT2,SkT3} and \textbf{Skxs20} satisfy all but one constraints of Ref. \cite{Dutra2012}. If one increases the admissible range of $\delta$ up to the interparticle distance, then two more parametrizations satisfying all the constraints of Ref. \cite{Dutra2012} and three more satisfying all but one will pass the $\delta$-correction test (see Table \ref{result}). At the same time, as seen in Fig. \ref{fdelt}, many parametrizations yield values of $\delta$ correction close to the range allowed by (\ref{res}). An interesting but not surprising observation is that, while some of the parametrizations of the family \textbf{SkT} pass the $\delta$ constraint test, others do not, although the parametrizations are based on the same inputs and use the same method. This observation may suggest that some of the parametrizations pass the test just by chance.

The main features of the parametrizations that passed the test are shown in (Tab. \ref{methods}, Appendix \ref{appC}). As seen from the Table, the families with acceptable values of the Tolman correction use finite nuclei properties connected with the nuclear surface (\textit{e.g.} surface properties of selected magic and semimagic nuclei, surface thickness, neutron rms radius) as input data. At the same time, families that did not pass the test use different input data (\textit{e.g.} \textbf{v070} or \textbf{skz2}). It should be also noted that parametrizations elaborated for neutron matter (\textit{e.g.} the \textbf{SLy} family) fail to produce an acceptable curvature correction to the surface tension in symmetric nuclear matter.

The obtained values appear negative for all chosen parametrizations except for \textbf{SV} (see Fig. \ref{fdelt}), which means that the surface of tension is located closer to the liquid phase with respect to the equimolar surface.

An important observation is that all parametrizations constructed by systematic variations of the parameters (e.g. \textbf{SV} and \textbf{SkS} families) do pass the $\delta$-correction test.

Attempting to find an apparent pattern of the forces performance in describing properties of the nuclear surfaces, we analyzed the dependence of $\delta$ on various force parameters and the macroscopic properties characteristics for the tested forces. We also calculated the coupling constants found in recent years and aimed to improve the relevance of the Skyrme forces to different physical properties \cite{Dutra2012} and given as linear combinations of individual parameters
\begin{equation}
\label{coupconst1}
 \begin{array}{l}
   {C_0}^\rho=\frac{3}{8}t_0+\frac{3}{48}t_3{\rho_0}^\alpha\\
   {C_1}^\rho=-\frac{1}{4}t_0\left(\frac{1}{2}+x_0\right)-\frac{1}{24}t_3\left(\frac{1}{2}+x_3\right){\rho_0}^\alpha.
 \end{array}
\end{equation}
Other combinations of the individual parameters are those providing the most compact formulation of the energy functional and the residual interaction \cite{Reinhard1999}
\begin{equation}
\label{coupconst2}
 \begin{array}{l}
   {b_0}=t_0\left(1+\frac{1}{2}x_0\right),\\
   {b_3}=\frac{1}{4}t_3\left(1+\frac{1}{2}x_3\right).
 \end{array}
\end{equation}

\begin{figure}[htp]
\includegraphics[scale=1.0]{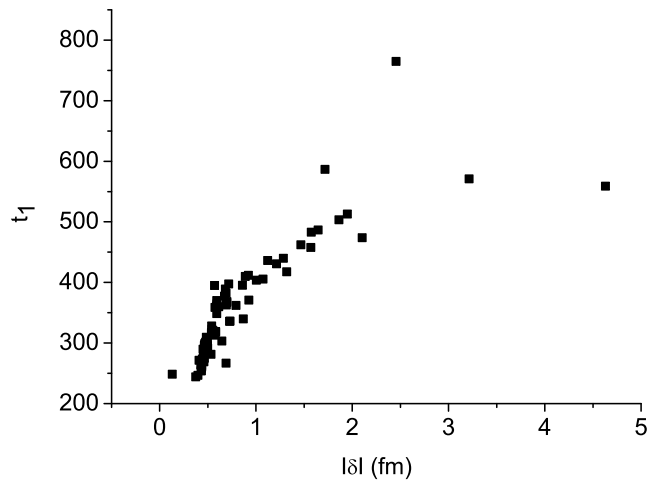}
\caption{Dependence of absolute value of $\delta$ on $t_1$ coefficient of the Skyrme force} \label{tone}
\end{figure}

\begin{figure}[htp]
\includegraphics[scale=1.0]{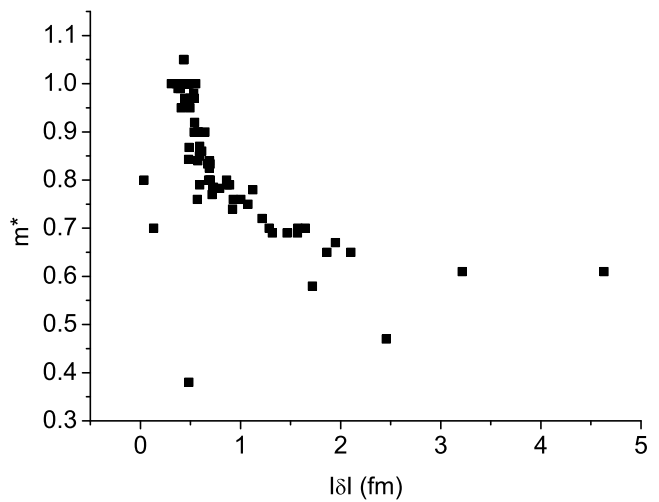}
\caption{Dependence of absolute value of $\delta$ on effective mass} \label{meff}
\end{figure}

\begin{figure}[htp]
\includegraphics[scale=1.0]{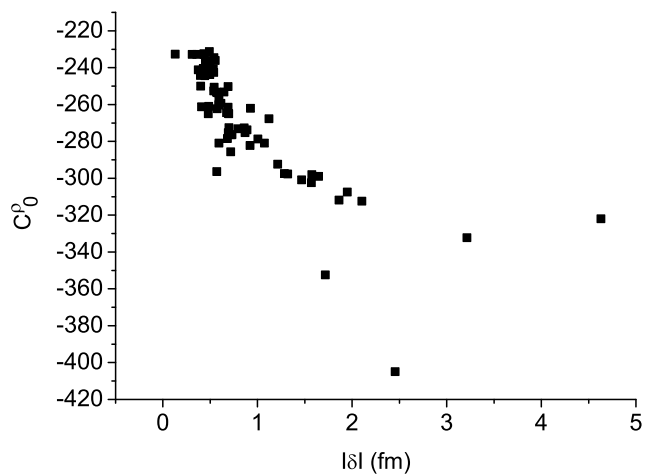}
\caption{Dependence of absolute value of $\delta$ on ${C_0}^\rho$} \label{Cro0}
\end{figure}

Unfortunately, no particular dependence of the value of $\delta$-correction in almost all of the force parameters, coupling constants or macroscopic properties was found. The only observable correlations are in the slight increase of the absolute value of $\delta$ with increasing $t_1$ (Fig. \ref{tone}), decreasing effective mass (Fig. \ref{meff}) and increasing absolute value of the coefficient ${C_0}^\rho$ (Fig. \ref{Cro0}).
From standard statistical analysis of the forces that pass the curvature correction test one may suggest that with probability 0.95 the most probable values of the above parameters should be
\begin{equation}
\label{avC0}
 \begin{array}{l}
   {C_0}^\rho=-245.1\pm3.6\\
   {m*}=0.96\pm0.02\\
   {t_1}=304.1\pm10.2\\
  \end{array}
\end{equation}
The standard deviations are $SD_{{C_0}^\rho}=13.8$, $SD_{m*}=0.07$ and $SD_{t_1}=36.7$.
One can see from the pictures (\ref{tone})-(\ref{Cro0}) that there exist a dependence of $\delta$ on $t_1$, $m*$ and ${C_0}^\rho$. Even though this may be the necessary condition for the force ability to describe nuclear surfaces, certainly it is not sufficient, since some of the  parametrizations, including those  with $\delta$ outside our range (\ref{res}), produce $t_1$, $m*$ and ${C_0}^\rho$ within the ranges imposed by Eq. (\ref{avC0}). This may be an indication of some basic problem in the Skyrme-type parametrizations, maybe connected with the large freedom in choosing the force parameters.

\section{Conclusions.}
In this paper we calculated the curvature correction term of the surface tension from the nuclear equation of state for 97 different standard Skyrme parametrizations available in the literature. The obtained results show strong dependence of the curvature correction term on the EOS.

To summarize, our study shows that not all the existing parametrizations are capable of describing adequately the interphase interfaces in nuclear matter. In spite of the considerable uncertainty regarding the absolute value of $\delta$, the suggested constraint, even with a wide admissible range for  $\delta$, allows one to test various Skyrme forces regarding their capacity to describe curvature effects. It should be mentioned that better agreement is observed with the use of EOS that account for the surface effects with respect to those that do not (or were designed for neutron matter).

Comparison of the obtained results with the available data on different Skyrme forces with nuclear matter constraints suggests two parametrization, namely \textbf{Ska25s20} and \textbf{SV-min}, that satisfy all constraints. For an extended admissible range for $\delta$ that number increases and two more parametrizations, namely \textbf{LNS} and \textbf{SCMC700}, come into play. It can be seen from our study that systematic variation of the parameters is quite efficient, giving promising results in studies of the basic properties of nuclear matter. At the same time, further progress with Skyrme type forces requires a better understanding of the physical  meaning of different parameters rather than the introduction of new parameters.

\begin{acknowledgments}
The work was supported by the National Natural Science Foundation of China under Grants No. 11161130520 and 11025524, the National Basic Research Program of China under Grant No. 2010CB832903, and the European Commissions 7th Framework Programme (FP7-PEOPLE-2010-IRSES) under Grant Agreement Project No. 269131. L.J. was supported by the grant {\textquotedblleft}Matter under extreme conditions{\textquotedblright} of the National Academy of Sciences of Ukraine
\end{acknowledgments}

\appendix

\section{\label{appA} Parametrizations used}
The Skyrme parametrizations analyzed in the work are
 BSk1 \cite{Samyn2002}, BSk10 \cite{Goriely2006}, E \cite{Friedrich1986},
 E$\sigma$ \cite{Friedrich1986}, FPLyon \cite{Meyer1993},
 G$\sigma$ \cite{Friedrich1986}, KDE \cite{Agrawal2005}, KDE0v \cite{Agrawal2005},
  KDE0v1 \cite{Agrawal2005}, LNS \cite{Cao2006}, MSk1 \cite{Tondeur2000},
 MSk2 \cite{Tondeur2000}, MSk3 \cite{Tondeur2000}, MSk4 \cite{Tondeur2000},
 MSk5 \cite{Tondeur2000}, MSk5* \cite{Farine2001}, MSk6 \cite{Tondeur2000},
 MSkA \cite{Sharma1995}, MSL0 \cite{Chen2010},
 NRAPR \cite{Steiner2005}, PRC45 \cite{Lee2001}, RATP \cite{Rayet1982},
 R$\sigma$ \cite{Friedrich1986}, SGII \cite{Giai1981}, SGO2 \cite{Shen2009},
 SII \cite{Beiner1975}, SIII \cite{Beiner1975},
 SIV \cite{Beiner1975}, SK255 \cite{Agrawal2003}, SK272 \cite{Agrawal2003},
 Ska \cite{Kohler1976}, Ska25s20 \cite{Dutra2012}, Ska35s20 \cite{Dutra2012},
 SkI4 \cite{Reinhard1999}, SkM* \cite{Chabanat1997}, SkMP \cite{Shen2009},
 SkO \cite{Reinhard1999}, SkP \cite{Chabanat1997},
 SKRA \cite{Rashdan2000}, SkS1 \cite{Gomez1992}, SkS2 \cite{Gomez1992},
 SkS3 \cite{Gomez1992}, SkS4 \cite{Gomez1992}, SkSC1 \cite{Pearson1991},
 SkSC10 \cite{Onsi1994}, SkSC2 \cite{Pearson1991}, SkSC3 \cite{Pearson1991},
 SkSC4 \cite{Onsi1994}, SkSC5 \cite{Onsi1994}, SkSC6 \cite{Onsi1994},
 SkSP1 \cite{Farine2001}, SkT1 \cite{Tondeur1984}, SkT1* \cite{Tondeur1984},
 SkT2 \cite{Tondeur1984}, SkT3 \cite{Tondeur1984}, SkT3* \cite{Tondeur1984},
 SkT4 \cite{Tondeur1984}, SkT5 \cite{Tondeur1984}, SkT6 \cite{Tondeur1984},
 SkT7 \cite{Tondeur1984}, SkT8 \cite{Tondeur1984}, SkT9 \cite{Tondeur1984},
 SKX \cite{Brown1998}, SKXce \cite{Brown1998}, SKXm \cite{Brown1998},
 Skxs15 \cite{Brown2007}, Skxs20 \cite{Brown2007}, Skxs25 \cite{Brown2007},
 Skz2 \cite{Margueron2002}, SLy4 \cite{Chabanat1998}, SLy5 \cite{Chabanat1998},
 SLy6 \cite{Chabanat1998}, SLy7 \cite{Chabanat1998}, SQMC650 \cite{Dutra2012},
 SQMC700 \cite{Dutra2012}, SV \cite{Beiner1975}, SV-bas \cite{Klupfel2009},
 SVI \cite{Beiner1975}, SV-K226 \cite{Klupfel2009}, SV-kap02 \cite{Klupfel2009},
 SV-mas08 \cite{Klupfel2009}, SV-min \cite{Klupfel2009}, SV-sym32 \cite{Klupfel2009},
 SV-tls \cite{Klupfel2009}, T \cite{Friedrich1986}, v070 \cite{Pearson2001},
 Z \cite{Friedrich1986}, ZR1a \cite{Lee2001}, ZR1b \cite{Lee2001}, ZR1c \cite{Lee2001},
 ZR2a \cite{Lee2001}, ZR2b \cite{Lee2001}, ZR2c \cite{Lee2001}, ZR3a \cite{Lee2001},
 ZR3b \cite{Lee2001}, ZR3c \cite{Lee2001}, Z$\sigma$ \cite{Friedrich1986},

\section{\label{appB} Equation of state of nuclear matter}
To calculate the $\delta$-correction, an EOS of nuclear matter at low-temperatures and in the high-densities limit was used, where $\lambda^3\rho\gg1$ ({\it i.e.} when the average de Broglie thermal wavelength $\lambda$ is larger than the average interparticle separation $\rho^{-\frac{1}{3}}$). In this case, the EoS takes the form \cite{Lee2008}:
\begin{equation}
\label{eos1}
  \begin{array}{l}
   P(\rho_q,T)=\sum\limits_{q}^{\;}\left[\frac{5}{3}{\varepsilon^*}_{kq}(\rho_q,T)-{\varepsilon}_{kq}(\rho_q,T)\right]\\
   +\frac{t_0}{2}\left(1+\frac{x_0}{2}\right)\rho^2+\frac{t_3}{12}\left(1+\frac{x_3}{2}\right)(\alpha+1)\rho^{\alpha+2}\\
   -\frac{t_0}{2}\left(x_0+\frac{1}{2}\right)\sum\limits_{q}^{\;}{\rho_q}^2-\frac{t_3}{12}\left(\frac{1}{2}+x_3\right)(\alpha+1)\rho^\alpha\sum\limits_{q}^{\;}{\rho_q}^2,
  \end{array}
\end{equation}
with
\begin{equation}
\label{eos2}
  \begin{array}{l}
   \varepsilon_{kq}=\frac{{m^*}_q}{m}\frac{1}{\beta}\frac{2g}{\sqrt{\pi}}{\lambda_q}^{-3}F_\frac{3}{2}(\eta_q),\\
   {\varepsilon^*}_{kq}=\frac{1}{\beta}\frac{2g}{\sqrt{\pi}}{\lambda_q}^{-3}F_\frac{3}{2}(\eta_q),
   \end{array}
\end{equation}
where $m$ and $m^*$ are the mass and effective mass respectively, $T$ and $\rho$ are temperature and density, $q$ is the particle type ($q$=proton, neutron), $F$ is the Fermi integral, $\lambda=\sqrt{\frac{2\pi\hbar^2}{m^*T}}$ is the average de Broglie thermal wavelength, $g=2$ is the spin degeneracy factor, $t_0$, $t_3$, $x_0$, $x_3$ and $\alpha$ are the Skyrme force parameters and $\beta=\frac{1}{T}$.

\section{\label{appC} Methods and input data used in construction of different Skyrme parametrizations}
\begin{table*}[htp]
 \caption{Comparison of the methods and input data used in construction of different Skyrme parametrizations}
 \label{methods}
 \begin{ruledtabular}
  \begin{tabular}{lll}
  \textrm{Skyrme forces}&\textrm{Method}&\textrm{Input data}\\
  \colrule
     KDE0v1 & Simulated annealing method & Ground state properties of normal and exotic nuclei:\\
     && binding energy, charge radii and spin-orbit splitting,\\
     && radii for $1d_{5/2}$ and $1f_{7/2}$ neutron orbits in $^{17}$O and $^{41}$Ca\\
     && breathing mode energies, critical density $\rho_{cr}$,\\
     && positive slope of the symmetry energy up to 3$\rho_0$,\\
     && enhancement factor associated with GDR, Landau parameter $G'_0$\\
     && \\
     LNS & Brueckner-Hartree-Fock & Nucleon effective mass in symmetric nuclear matter (SNM)\\
     & with 2- and 3-body forces & and asymmetric nuclear matter (ANM), energy per\\
     &(homogeneous matter)& particle in SNM and ANM as function of density and\\
     &Hartree-Fock& proton neutron asymmetry;\\
     &(finite matter) & constraint on the Landau parameter $G_0$,\\
     && surface properties of selected magic and semimagic nuclei,\\
     && spin-orbit splitting $p_{1/2}-p_{3/2}$ in $^{16}$O\\
     && \\
     SV family & Hartree-Fock & Properties of finite nuclei (energies, radii and surface thickness), \\
      &  & energies of giant resonances,\\
      &  & systematic variations of selected nuclear matter properties\\
      && \\
     MSkA &  Density dependent Hartree-Fock & Empirical binding energies and charge radii of the: \\
      &  & closed shell nuclei $^{16}$O, $^{40}$Ca, $^{90}$Zr, and $^{208}$Pb;\\
      &  & isotopes $^{116,124}$Sn and $^{214}$Pb; doubly closed $^{132}$Sn;\\
      && \\
     SK255, SK272& Hartree-Fock based random phase & Nuclear binding energies, charge radii, and neutron radii;\\
      &  approximation (RPA) approach & rms charge radius of $^{208}$Pb; incompressibility coefficient from\\
      &  & relativistic and nonrelativistic mean field based RPA calculations \\
      && \\
     SkS family & Hartree-Fock seniority& Symmetric nuclear matter ground state properties;\\
      &  & binding energies of $^{16}$O, $^{40,48}$Ca, $^{90}$Zr, and $^{208}$Pb;\\
      &  & spin-orbit splitting in $^{16}$O; surface parameter and \\
      &  & symmetry properties; fission barrier of $^{240}$Pu; \\
      &  & restrictions on Landau parameter for \textit{SkS4} \\
      && \\
     SkT family & Hartree-Fock; & Nuclear radii and binding energies; droplet model parameters;\\
      & Extended-Thomas-Fermi & experimental masses, charge radii, charge distribution, neutron\\
      & & skin  thickness, semiclassical fission barriers and \\
      & & Landau parameters \\
      && \\
     Skxs family & Skyrme Hartree-Fock;& Binding energies, rms charge radii, and single-particle energies;\\
      & relativistic mean-field models & binding energy difference $^{48}$Ni-$^{48}$Ca; charge density of $^{208}$Pb;\\
      &  & constraint of $\alpha$=1/6 for the density dependent potential\\
      && \\
     E, Z, T, E$_\sigma$, G$_\sigma$, R$_\sigma$, Z$_\sigma$ & Hartree-Fock-Bogoliubov & Binding energy, diffraction radius, \\
      &  & surface width of $^{16}$O, $^{40,48}$Ca, $^{58}$Ni, $^{90}$Zr, $^{116,124}$Sn and $^{208}$Pb;\\
      &  & $l-s$ splitting of the $1p$ level in $^{16}$O\\
  \end{tabular}
 \end{ruledtabular}
\end{table*}

\bibliography{DeltBase2}

\end{document}